\documentclass{aa}
\usepackage{graphicx}
\usepackage{lscape}
\usepackage{subfigure}
\usepackage{natbib}
\bibpunct{(}{)}{;}{a}{}{,} 

\usepackage{txfonts}
\begin{document}
\title{An unusual very low-mass high-amplitude pre-main sequence periodic variable.}

 	\author{Mar\'{i}a V. Rodr\'{i}guez-Ledesma
          \inst{1,}\inst{2},         
          Reinhard Mundt\inst{1},
Mansur Ibrahimov
\inst{3},
Sergio Messina
\inst{4},
Padmakar Parihar
\inst{5},
Frederic V. Hessman
\inst{2},
Catarina Alves de Oliveira
\inst{6},
\and
William Herbst
          \inst{7}
	 }
\institute{Max-Planck-Institut f\"ur Astronomie (MPIA), K\"onigstuhl 17, 69117 Heidelberg, Germany\\
              \email{vicrodriguez@mpia.de}
	     \and
Institut f\"ur Astrophysik, Georg-August-Universit\"at, Friedrich-Hund-Platz 1, 37077 G\"ottingen, Germany\\
\and
Ulugh Bek Astronomical Institute of the Uzbek Academy of Sciences, Astronomicheskaya 33,
Tashkent 700052, Uzbekistan\\ 
\and
INAF-Osservatorio Astrofisico di Catania, via S. Sofia 78, 95123 Catania, Italy\\ 
\and
Indian Institute of Astrophysics, Bangalore 560034, India\\
\and
Herschel Science Centre, European Space Astronomy Centre (ESA), P.O. Box, 78, 28691 Villanueva de la Ca\~{n}ada, Madrid, Spain\\
\and
Astronomy Department, Wesleyan University, Middletown, CT 06459 USA }


 
  \abstract
   {}
{We have investigated the nature of the variability of CHS\,7797, an unusual periodic variable in the Orion Nebula Cluster.}
{An extensive I-band photometric data set of CHS\,7797 was compiled between 2004-2010 using various telescopes. 
Further optical data have been collected in R and z$\arcmin$ bands. In addition, simultaneous observations of the ONC region including CHS\,7797 were performed in the I, J, K$_{s}$ \& IRAC $3.6$ and $4.5\,\mu m$ bands over a time interval of $\approx$\,40\,d. }
{CHS\,7797 shows an unusual large-amplitude variation of $\approx$\,1.7 mag in the R, I, and z$\arcmin$ bands with a period $17.786\,\pm\,0.03$\,d ($FAP\,=\,1x10^{-15}\%$).
The amplitude of the brightness modulation decreases only slightly at longer wavelengths. 
The star is faint during $\approx$\,2/3 of the period and the shape of the phased light-curves for the seven different observing seasons shows minor changes and small-amplitude variations. 
Interestingly, there are no significant colour-flux correlations for $\lambda$\,$\lesssim$\,2\,$\mu$m, while the object becomes redder when fainter at longer wavelengths.
CHS\,7797 has a spectral type of M6 and an estimated mass between 0.04-0.1\,M$_{\odot}$.
}
{The analysis of the data suggests that the periodic variability of CHS\,7797 is most probably caused by an orbital motion. 
Variability as a result of rotational brightness modulation by a hot spot is excluded by the lack of any colour-brightness correlation in the optical. 
The latter indicates that CHS\,7797 is most probably occulted by circumstellar matter in which grains have grown from typical 0.1\,$\mu$m to $\approx$\,1-2\,$\mu$m sizes. 
We discuss two possible scenarios in which CHS\,7797 is periodically eclipsed by structures in a disc, namely that CHS\,7797 is a single object with a circumstellar disc, or that CHS\,7797 is a binary system, similar to KH\,15D, in which an inclined circumbinary disc is responsible of the variability. 
Possible reasons for the typical 0.3\,mag variations in I-band at a given phase are discussed.
}
\titlerunning{A very low-mass KH\,15D-like system in the ONC? }
\authorrunning{Rodr\'iguez-Ledesma et al.} 
\maketitle
\section{Introduction}

\begin{figure*}
\includegraphics[scale=0.65]{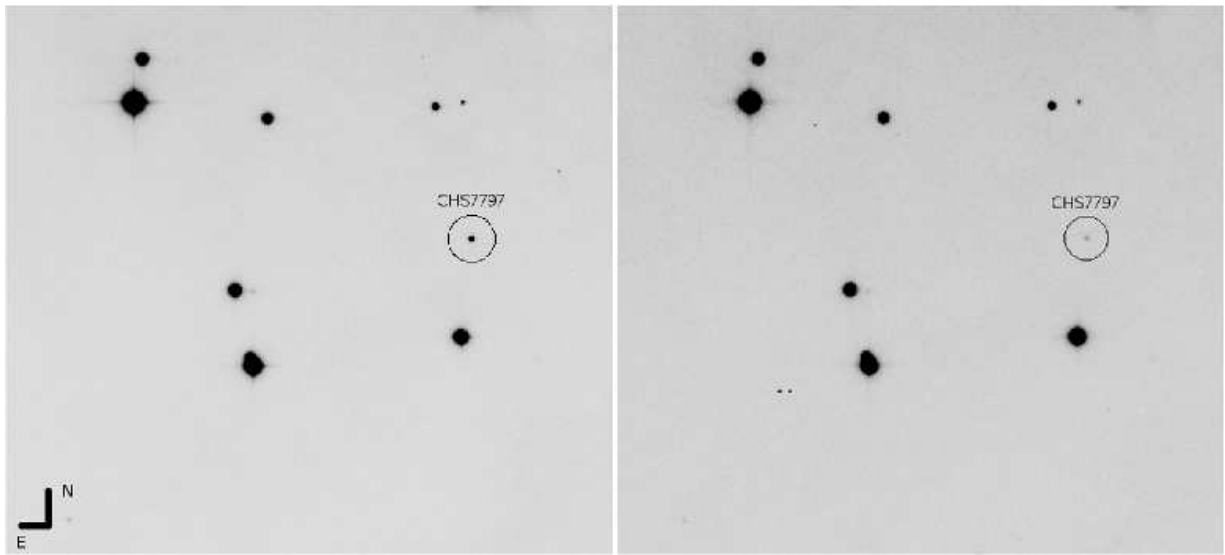}
\caption{I-band images of CHS\,7797 during maximum and minimum brightness obtained with WFI at the 2.2\,m telescope on La Silla, Chile. North is up and east is left. The region shown is 2\,$\arcmin$x2\,$\arcmin$ in size. }
\label{Fig_FC}
\end{figure*}

The photometric variability of T Tauri stars (TTS) is attributed in many cases to magnetically induced cool starspots and/or magnetically channelled variable accretion flows, which produce hot spots at the base of the magnetic channels. During the last 10-15 years, various extensive CCD monitoring programs of young clusters like NGC\,2264 and the Orion Nebular Cluster (ONC) have been carried out. These monitoring programs have significantly increased our knowledge of TTS variability and rotation, and have revealed thousands of mostly low-amplitude periodic pre-main sequence (PMS) variables and a similar number of irregular variable PMS stars \citep[e.g.][]{Herbst2002, Lamm2004, Rodriguez-Ledesma2009}. As reviewed by \cite{Herbst2007}, and based on many of these studies, it is possible to distinguish five types of PMS variability. The different types have known properties such as typical amplitude range of the variability at different wavelengths. For example, amplitudes at optical wavelengths of periodic brightness modulation, due to hot spots and/or magnetically channelled accretion, can be a factor 2 to 5 larger than the periodic variations believed to be caused by cool spots, which cannot be larger than 0.5 mag in I \citep{Herbst1994}. Another relevant difference between these two cases is that cool spots have been observed to be stable over thousands of cycles, while accretion-related variability is much more unstable and irregular, with the observed periodicity often only surviving several tens of cycles. It is important to note that rotational properties and accretion-related variability have been observed also in substellar objects \cite[e.g.][]{Scholz2004,Mohanty2005,Rodriguez-Ledesma2009}.

Because most of these variability studies involve a large number of objects, they are not only valuable for characterising variability, but also for detecting interesting and rare objects. One striking example is KH\,15D, a unique 48\,d periodic variable in NGC\,2264, with extremely deep ($\approx$3.5 mag) minima. It was first recognized as interesting by \cite{Kearns1998}. \cite{Hamilton2001} reported that KH\,15D is a K7 PMS member of NGC\,2264 ($\approx$ 2-4\,Myr), with a mass of 0.5-1\,$M_{\odot}$, and together with \cite{Herbst2002} proposed that the star was being eclipsed by circumstellar material. Further intense investigations and, in particular, radial velocity studies led to a binary model for KH\,15D in which the system is surrounded by a nearly edge-on and slowly precessing circumbinary disc, to which the binary system is slightly inclined \citep[and references therein]{Winn2006, Herbst2008}. Over the past 15 years the circumbinary disc has completely occulted the orbit of component B and has allowed us to see only a diminishing fraction of the orbit of star A. It is the appearing and disappearing
of star A behind the disc that causes the strong light modulation. 

Other TTS binaries are expected to have similar 
characteristics, but KH15\,D is not only special because of its particular geometric orientation and the precessing circumbinary disc but also because of the indications of significant grain growth within the disc \citep{Herbst2008}. This binary is also interesting because it is a jet-driving source, with the jet being most probably a ``common product`` of the whole binary system or the circumbinary disc \citep{Mundt2010}. Although thousands of PMS stars have been photometrically monitored in the past years, only few objects show a variable behaviour that resembles KH\,15D, i.e. YLW\,16A and WL\,4 in $\rho$\,Oph \citep{Plavchan2010,Plavchan2008}, and V718\,Per in IC 348 \citep{Grinin2008}. In the case of WL\,4 and YLW\,16A, the authors proposed the existence of a new class of periodic disc-eclipsing binaries to explain the KH\,15D type variability, in which a third body in these systems is responsible for the warping of the inner circumbinary disc needed to produce the eclipses, while \cite{Grinin2008} argued that V718\,Per is most probably a single star surrounded by an edge-on circumstellar disc with an irregular mass distribution at the inner edge of the disc, which causes the observed periodic variability. AATau-like objects show also high amplitude variability, although this variability is found to be only quasi-periodic \citep{Alencar2010}. As described in \cite{Alencar2010}, in the fairly common AATau-like objects the quasi-periodic variability is attributed to a magnetically controlled inner disc warp.

\cite{Rodriguez-Ledesma2009} carried out a photometric monitoring in the ONC during 2004 with the primary goal of deriving rotational periods for a large sample of very low-mass PMS stars and brown dwarfs. Several of the objects in this sample show unusual light curves: CHS\,7797 (i.e. star No.7797 in \cite{Carpenter2001}) stuck out particularly because of its unusual large-amplitude modulation of $\approx$1.7\,mag in I and its periodic variability.
Hoping that this newly discovered periodic variable might have a similar potential to KH\,15D, which has provided considerable insight into the physics of PMS binary systems and their circumbinary discs, we have carried out a photometric follow-up campaign, including multicolour data collected with various telescopes/instruments.

This paper is organised as follows: in Section\,2, we present the optical and near-infrared (NIR) data sets, while in Section\,3 we describe the photometric data evaluation. Section\,4 describes the time series analysis procedure. We analyse colour changes in Section\,\ref{Colours} and estimate luminosity and mass ranges for CHS\,7797 in Section\,\ref{Luminosity}. A detailed discussion of our results is given in Section\,\ref{Discussion}, while final conclusions are presented in Section 7. 

\section{Optical and near-infrared data set}

CHS\,7797 ($RA(2000)=05:35:08.7$ \,\,$DEC(2000)=-05:31:27.3$) is located $\approx$\,9\,$\arcmin$ south-west of the centre of the ONC . Fig.\,\ref{Fig_FC} shows I-band images of CHS\,7797 taken at maximum and minimum brightness. 
\begin{figure}
\includegraphics[trim = 6mm 0mm 0mm 0mm,clip,width=8.9cm, height=7cm]{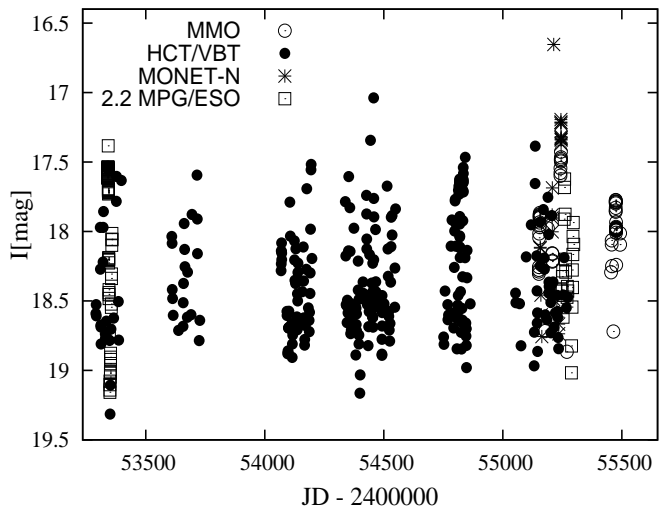}
\caption{I-band light-curves from the 2004 to the 2010 observing season. Observations from the different sites are shown with different symbols (see Table\,1 and text for details).}
\label{Fig_lightcurves}
\end{figure}
\begin{table}
\centering
\begin{minipage}[c]{\columnwidth}
\renewcommand{\arraystretch}{1.6}  
\caption{Overview on the obtained photometric observations.}
\label{Obs_nights}
\addtolength{\tabcolsep}{0.0pt}
{\Large
\resizebox{8.8cm}{!} {
\begin{tabular}{l|c|c|c} 
\hline\hline             
Site & Observing\footnote{Observing seasons are defined as follows:\\A=2004, B=2005, C=2006/2007, D=2007/2008, E=2008/2009, F=2009/2010, G=2010} &Bands & Number of\\
& seasons & &observations\\
\hline
La Silla 2.2m (WFI) & A,\,F & I & 106\\
Maidanak 1.5m & F,\,G& I,\,R,\,z$\arcmin$& 233\\
HCT and VBT (2 and 2.3\,m) & A-F& I,\,R& 346\\
MONET 1.2m & F& I& 24\\
UKIRT (WFCAM)& F& J& 59\\
CFHT (WIRCam)& F& K$_{s}$& 10 \\
Spitzer (IRAC) & F& $3.6\mu$m,\,$4.5\mu$m& 88\\
\hline
\end{tabular}
}
}
\renewcommand{\footnoterule}{} 
\end{minipage}
\end{table}
\cite{Rodriguez-Ledesma2009} were only able to derive an approximate period of 18\,-\,20\,d from their data.
This initial discovery triggered our photometric follow-up observations at various telescopes during the years 2009 and 2010. 
Observations were carried out from November 2009 to November 2010 with the 2.2m telescope on La\,Silla/Chile, the 1.2m MONET/N Telescope at McDonald Observatory/USA, and with the 1.5m telescope at Mount Maidanak Observatory (MMO) in Uzbekistan. In addition, we have also made use of the extensive archival data of the ONC (including the region around CHS\,7797) collected between 2004 and 2009 with the 2m Himalayan Chandra Telescope (HCT) at the Indian Astronomical Observatory and the 2.3m Vainu Bappu Telescope (VBT) at the Vainu Bappu Observatory, both in India \citep[see][for details]{Parihar2009}. The combination of all these data provides a unique I-band data set consisting of 663 observations over a total of seven observing seasons. Additional optical observations at other wavelengths were obtained at the MMO (R and z$\arcmin$) and from HCT and VBT (R and V).
In addition, J, H, K$_{s}$ images are available from observations with the CFHT Wide field InfraRed Camera (WIRCam) and the UKIRT Wide Field CAMera (WFCAM) carried out in 2006 and 2009, respectively (Alves de Oliveira, priv. comm.). We complemented our data set with recently released Spitzer/IRAC observation of CHS\,7797 using the IRAC1 (3.6$\mu$m) and IRAC2 (4.5$\mu$m) filters taken as part of the YSOVAR project during 2009 \citep{MoralesC2011}. Table\,1 provides an overview of the multiwavelength photometric observations on CHS\,7797. Very important for a detailed monitoring of CHS\,7797 were the Spitzer observations, the WFCam observations, the WIRCam/CFHT observations, and the I-band observations (MONET/N, MMO, and HCT) between October 23 and December 4 2009, which were taken at this overlapping time interval by chance (see Fig. \ref{Fig_lightcurves3}).
This rich multi-wavelength data set allows us to put constraints on the physical characteristics of CHS\,7797. Particularly for young and highly variable objects, such simultaneous data are of great importance for understanding their nature.

\begin{figure*}
\includegraphics[scale=1.4]{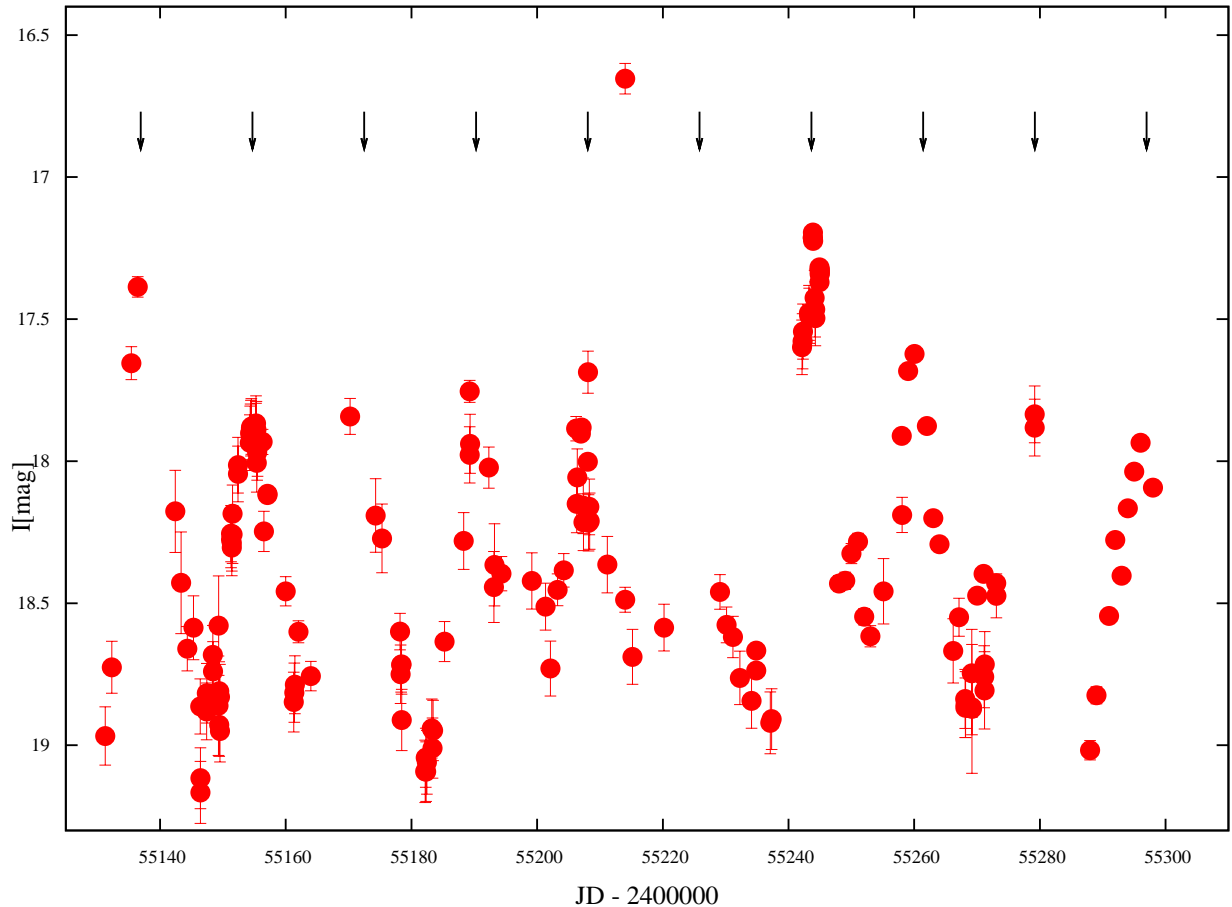}
\caption{I-band light-curves for the 2009/2010 observing season during which observations were performed almost continuously over $\approx$\,4 months. Arrows show the observed and expected positions of the maximum light. During these season a ``supermaximum'' was observed at JD\,=\,2455214 (see section\,\ref{Discussion} for details).}
  \label{Fig_lightcurves1}
   \end{figure*}
\section{Photometry}

For the observations taken with the 2.2m, MMO, and MONET/N, aperture photometry was performed using standard IRAF tasks. A set of known non-variable or very low variable stars close to CHS\,7797 were selected to perform relative photometry between these secondary standards and CHS\,7797. These non-variable stars were selected from the detailed monitoring of the field by \cite{Rodriguez-Ledesma2009}. With the MMO observations of Landolt standard star fields and SDSS images the absolute magnitudes of these secondary standards in R, I and z$\arcmin$ could be determined. A table with these magnitudes is given in the appendix. For the observations taken at HCT and VBT we refer the reader to \cite{Parihar2009}.

Fig.\,\ref{Fig_lightcurves} shows the I-band light-curve of the whole data set covering the seven observing seasons (i.e. from Feb. 2004 to Nov. 2010). The observations from the various sites are shown with different symbols. It is clear from Fig.\,\ref{Fig_lightcurves} that during 2007 and 2009 relatively dense observations over several months were carried out. Some variations in the amplitude of the brightness modulation can be seen over the six years. For example, during the first observing season $\approx$\,0.5\,mag fainter minima were observed while during the 2009/2010 observing season $\approx$\,0.3\,mag brighter phases were recorded. Fig.\,\ref{Fig_lightcurves1} shows the light-curve during the 2009/2010 observing season. The arrows show the expected maximum positions based on the period measured (see Section \ref{Period}), which agree very well with the observed maximum positions. Variations in the peak brightness by up to $\approx$\,0.5\,mag are evident from Fig. \ref{Fig_lightcurves1}. Very interesting is the enormous brightening of CHS\,7797 at JD\,2455214 with I\,$\approx$\,16.7\,mag, i.e. more than 1\,mag brighter than the typical peak values. We will discuss the possible nature of this unusual ``supermaximum'' in Section \ref{Discussion}.    
 
\begin{figure}
\centering
\includegraphics[trim = 3mm 0mm 0mm 0mm,clip,scale=0.6]{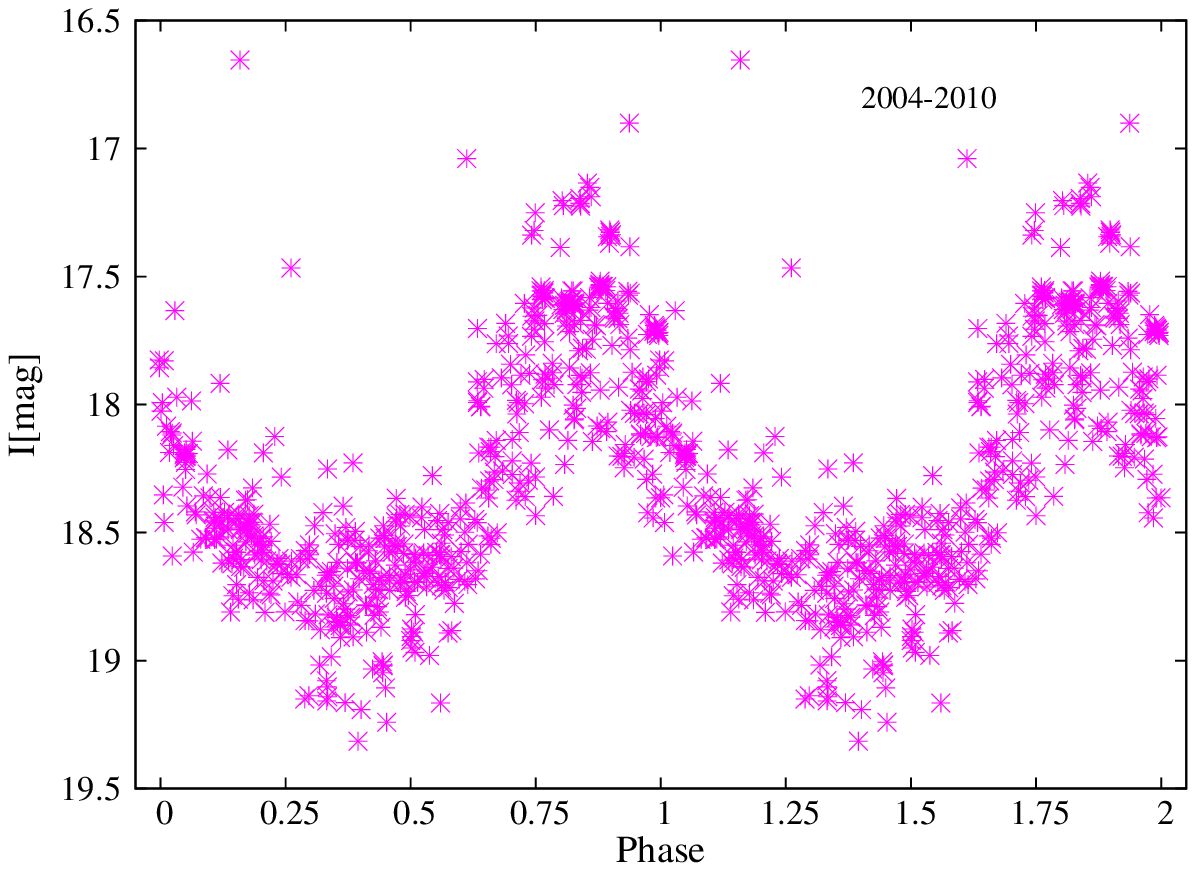}\\
\includegraphics[trim = 43mm 0mm 39mm 0mm,clip,scale=1.6]{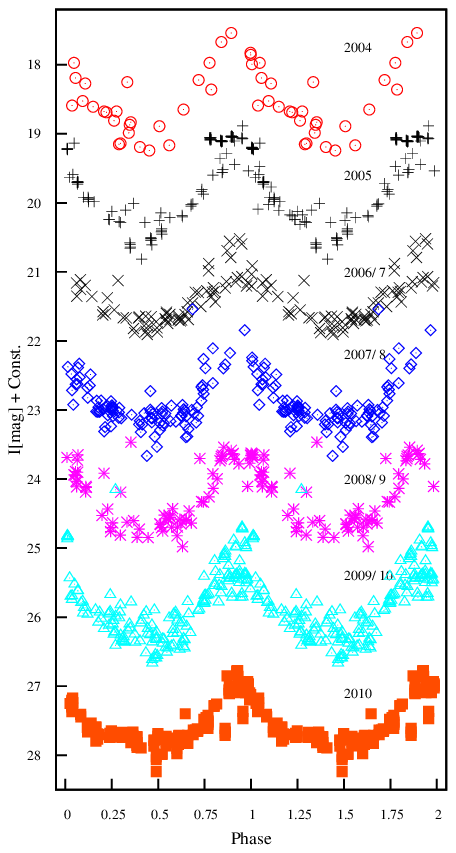}
\caption{\textit{Top:}\,Light-curve of the whole I-band data set from 2004 to 2010 phased with a period of 17.8\,d. \textit{Bottom:}\,Phased light-curves of the individual seven observing seasons shown with different symbols. Slight differences in the shape of these phased light-curves are visible.}
\label{Fig_phasedlc_season}
\end{figure}

\section{Period determination\label{Period}}

The period determination is based on the I-band observations, since they provide the by far most complete and dense data set, extending over a very long time base. In addition, young very low mass objects are brighter in this part of the spectrum and, particularly important, I-band images suffer much less from nebular background contamination than R-band images. 

To determine the periodicity of the light modulation we used a combination of two periodogram
analysis techniques: the Lomb-Scargle periodogram \citep{Scargle1982} and
the CLEAN algorithm \citep{Roberts1987}. As described in detail in \cite{Rodriguez-Ledesma2009}, these two techniques deliver a reliable period determination in which the effects of aliasing and false peaks due to inhomogeneous sampling no longer appear. 
We derived false-alarm probabilities (FAP) based on the Lomb-Scargle periodogram to test the significance of the period found. Independent of the Lomb-Scargle periodogram FAP calculation, we used the statistical F-test and a derived FAP from it, as described in \cite{Scholz2004a} and \cite{Rodriguez-Ledesma2009}. Since the FAP$_{F-test}$ represents the probability that the period found is caused by variations in the photometric noise, it is completely independent of the periodogram analysis. 

We determined a period for the individual seven observing seasons and for the whole data set from 2004 to 2010. The results agree within the errors in all cases. The derived period based on all 663 I-band observations is $P\,=\,17.786\,\pm\,0.03$\,days. The computed FAPs, based on the Scargle periodogram and the F-test, give extremely low probabilities ($\leq$1\,x\,10$^{-15}$) that the detected period is not real. Owing to the very long time base of our observations, we carefully searched in the periodograms for additional lower power peaks corresponding to longer periods, but no other significant peak was found in either the Scargle or the CLEAN periodogram. 

Fig.\,\ref{Fig_phasedlc_season} shows in the top panel the phased light-curve based on the whole I-band data set. The bottom panel of Fig.\,\ref{Fig_phasedlc_season} shows the phased light-curves of the individual seven observing seasons. Slight differences in the shape of the phased-light curves for different seasons are clearly seen. For example, the phased light-curves for the years 2004 and 2005 show pronounced maximum and minimum brightness phases (i.e. triangular shape), while during the seasons 2006/2007 and 2008/2009 the maxima of the phased light-curves are slightly flatter. It is also evident that there are some changes in the amplitude between seasons. A more detailed description and discussion of these features is given in Section\,\ref{Discussion}.

\section{Multiwavelength data}

\begin{figure}
\includegraphics[trim = 3mm 0mm 0mm 0mm,clip,width=8.9cm, height=6.5cm]{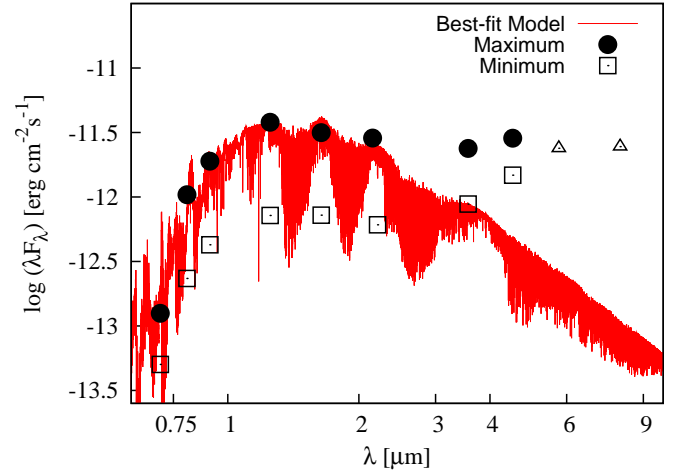}\\
\caption{Spectral energy distribution (SED) of CHS\,7797 during maximum (circles) and minimum (open squares) brightness. Open triangles show IRAC measurements at 5.6\,$\mu$m and 8\,$\mu$m, which are single-epoch observations taken about 3-4 days after maximum. The best-fitted model found by VOSA, i.e. NextGen models with T$_{eff}$\,=\,2700\,K and $log\,g\,=\,4$, is shown. Note that brightness difference between minimum and maximum decreases for $\lambda$\,$\gtrsim$ 2 $\mu$m.}
\label{Fig_SED}
\end{figure}

In addition to the optical data set, we analysed NIR observations taken during November 2009 in the J, K$_{s}$, IRAC1 (3.6$\mu$m) and IRAC2 (4.5$\mu$m) bands. Fortuitously, some of the K$_{s}$-band observations from WIRCam/CFHT, those in the J-band from WFCAM/UKIRT and I-band observations were taken simultaneously with the Spitzer observations. These particular observations are extremely valuable because they allow us to study colour changes of CHS\,7797 as a function of wavelength and the spectral energy distribution (SED). In Fig\,\ref{Fig_SED}, these SEDs\footnote{We note that we used the simultaneous data for the I, J, K$_{s}$, IRAC1[$3.6\mu$m], and IRAC2[$4.6\mu$m] bands along with the R, z$\arcmin$, H, IRAC3[$5.8\mu$m], and IRAC4[$8.0\mu$m] data, which were taken at different epochs.} indicate that CHS\,7797 is a very red and embedded pre-main sequence star. If we take into account the classification based on the IR slope by \cite{Lada2006}, with a slope of 0.8, CHS\,7797 classifies as a Class\,I object. As evident from Fig.\,\ref{Fig_SED} the brightness difference between maximum and minimum light is nearly constant until 2.2\,$\mu$m, and for the longer wavelengths the difference is smaller. 

\begin{figure}
\includegraphics[width=8.8cm, height=7cm]{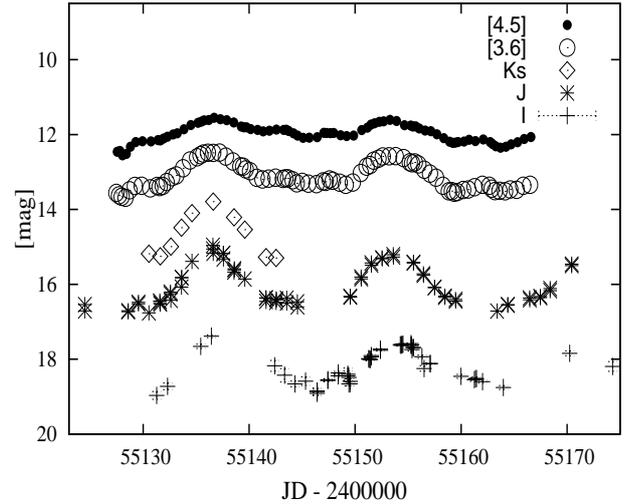}
\caption{Light-curves in I, J, K$_{s}$, $3.6\mu$m and $4.5\mu$m bands during part of the 2009/2010 observing season.}
\label{Fig_lightcurves3}
\end{figure}

The time interval over which we have obtained simultaneous observations in I, J, K$_{s}$, $3.6$ and $4.5\mu$m is displayed in Fig.\,\ref{Fig_lightcurves3}. This time interval of $\approx$\,40\,d covers more than two full period cycles and is best sampled by the IRAC data. The ptp amplitude show a slight decrease towards longer wavelength. The relative decrease 
is lower (20\%) in the optical/NIR from 1.7 in I to 1.4 in K$_{s}$\textbf{\footnote{Note from Table\,\ref{properties} that the ptp amplitudes in I and z$\arcmin$ are the same and equal to 1.7\,mag.}}, than in the IRAC bands (35\%), from 0.9 in 
$3.6\mu$ and to 0.6 in $4.5\mu$. Typical amplitudes in the observed bands are listed in Table\,\ref{properties}. Furthermore, the shape of the light-curves between the I and IRAC bands clearly looks different. While in the former case the light-curve is more sinusoidal with a pronounced minimum, the light-curves at the latter two wavelengths look more like brightness maxima superimposed on a roughly constant light level, resulting in practically flat and broad minima.
\subsection{Colour changes\label{Colours}}

\begin{figure*}
\includegraphics[width=8.8cm, height=6.5cm]{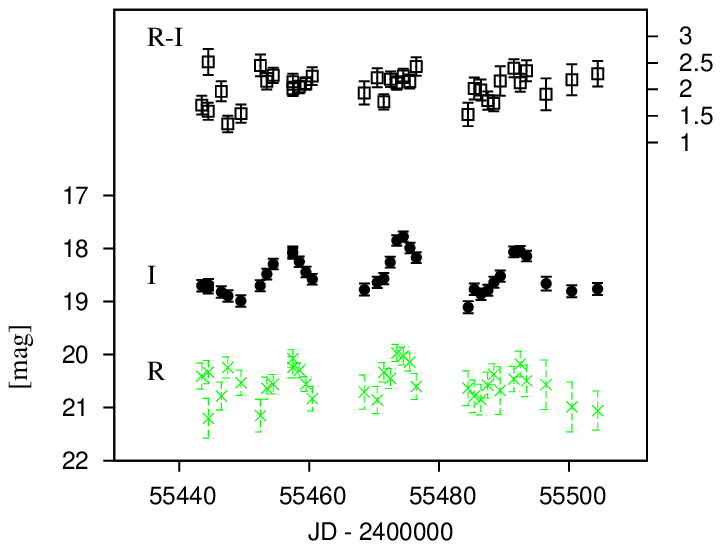}
\includegraphics[width=8.8cm, height=6.5cm]{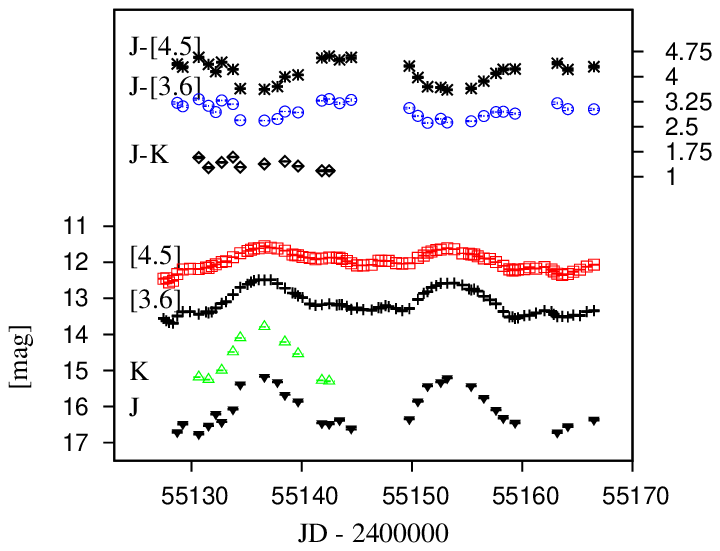}\\
\includegraphics[width=8.8cm, height=6.5cm]{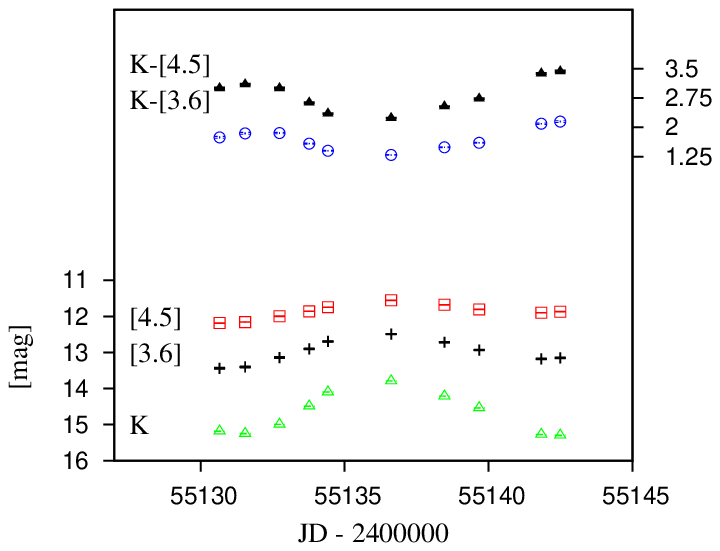}
\includegraphics[width=8.8cm, height=6.5cm]{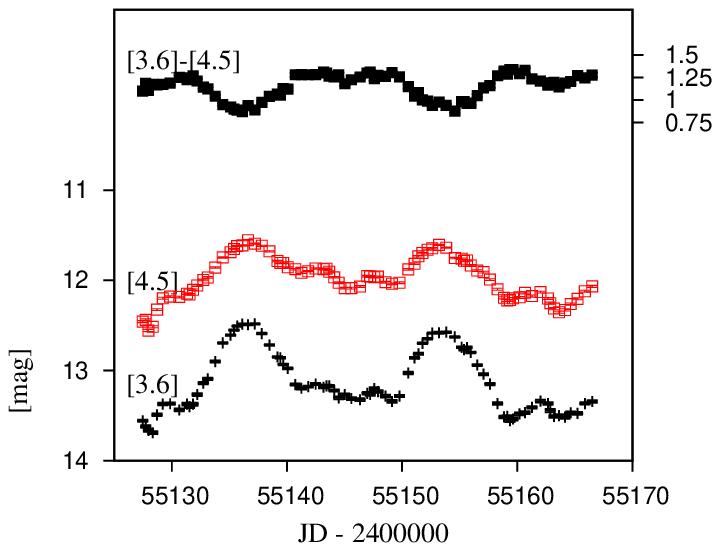}\\
\caption{Colour and magnitude variations for two different time intervals. The top-left panel shows the variations in R-I, I, and R. The other three panels show colours and magnitudes for wavelengths up to 4.5\,$\mu$m. No colour variations are evident for the R-I and J-K colours, while strong variations are visible for $\lambda$\,$\gtrsim$\,2\,$\mu$m. In these later cases the star becomes redder when becoming fainter (see also Fig.\,\ref{Fig_KK45}).}
\label{Fig_colour_nir}
\end{figure*}

To study the colour variations as a function of wavelength in more detail, we display several colour and magnitude light-curves in Fig.\,\ref{Fig_colour_nir}. Although variability is clear for all photometric bands, not all colours displayed in Fig.\,\ref{Fig_colour_nir} show variations that correlate with the magnitude variations, i.e. the object is not always becoming redder or bluer at bright or faint states. 

\begin{figure}
\includegraphics[width=8.7cm, height=6.2cm]{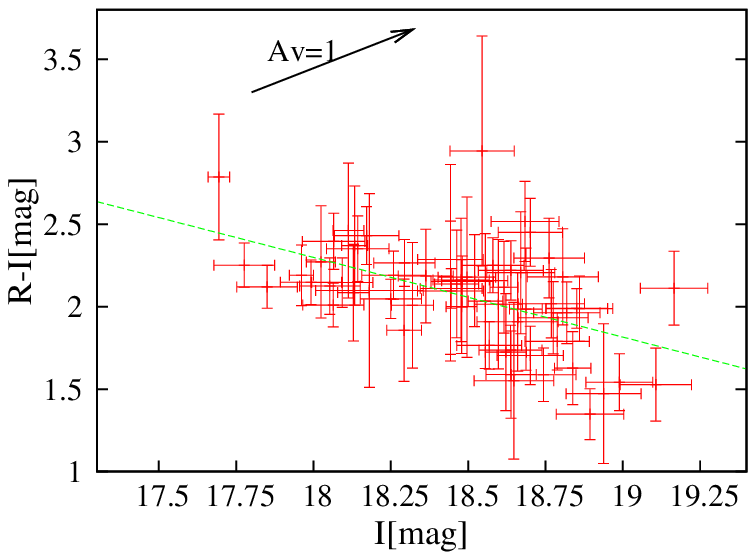}\\
\includegraphics[width=8.7cm, height=6.2cm]{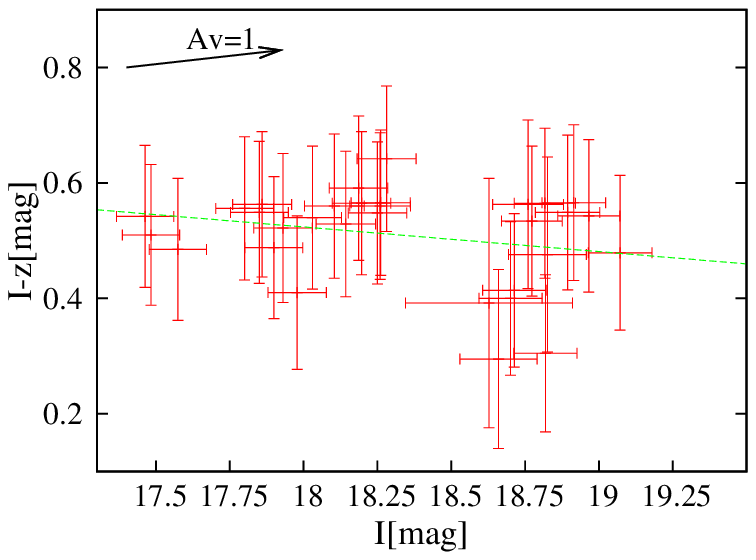}\\
\includegraphics[width=8.7cm, height=6.2cm]{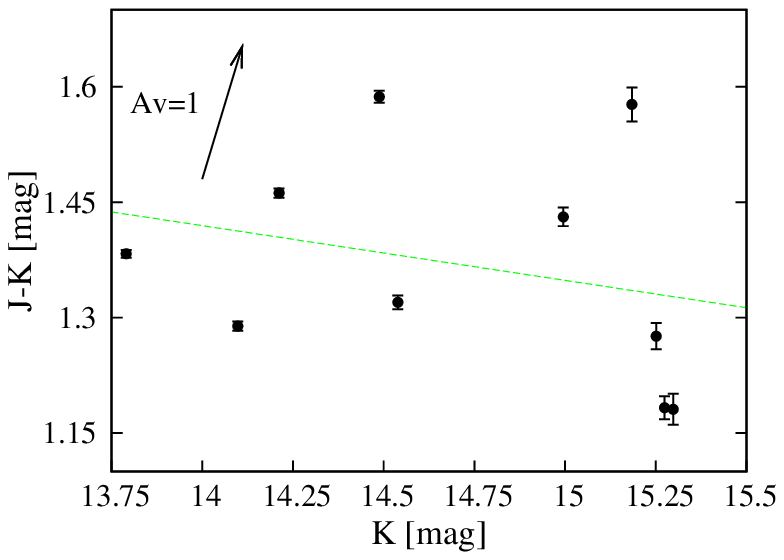}
\caption{Colour-magnitude correlations for (R-I) versus I, (I-z$\arcmin$) versus I, and (J-K$_{s}$) versus K$_{s}$, from top to bottom, respectively. An extinction vector for A$_v$=1 is plotted in all cases, indicating that any variation due to variable extinction by dust particles similar to those found in the interstellar medium would cause a quite different colour magnitude correlation. The dotted lines show the least-squares fit to the data, which is not statistically significant in any of the three correlations.}
\label{Fig_colourI}
\end{figure}

\begin{figure}
\includegraphics[width=8.7cm, height=6.3cm]{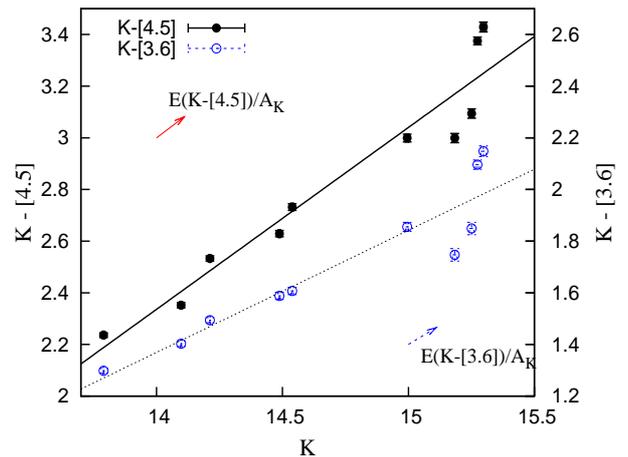}\\
\caption{Colour-magnitude correlations for K$_{s}$-$4.5\mu$m and K$_{s}$-$3.6\mu$m versus K$_{s}$. The lines show the linear fit to each correlation. In both cases the correlations are significant at a 3\,$\sigma$ level. Extinction vectors corresponding to A$_v$=1 are shown for comparison. The positive slope in both correlations agrees with variability caused by variations in extinction by dust particles smaller than the observing wavelength.}
\label{Fig_KK45}
\end{figure}

To gain a better understanding of the colour-brightness relations in Fig.\,\ref{Fig_colour_nir}, we display the following three correlations in Fig.\,\ref{Fig_colourI}: (R-I) versus I, (I-z$\arcmin$) versus I, and (J-K$_{s}$) versus K. Moreover, to study such correlation at longer wavelengths we show in Fig.\,\ref{Fig_KK45} the (K$_{s}$-$3.6\mu$m) and (K$_{s}$-$4.5\mu$m) colours as a function of the K$_{s}$ band magnitude.  

As evident from Fig.\,\ref{Fig_colourI}, there is no correlation between colours and brightness (correlation coefficient R$\approx$\,0.2 in all cases), quite in contrast to Fig.\,\ref{Fig_KK45}, where a clear and statistically significant correlation is observed between the K$_{s}$-$3.6\mu$m and K$_{s}$-$4.5\mu$m colours as a function of K$_{s}$ magnitude (R\,=\,0.97). As addressed by \cite{Carpenter2001} and \cite{Alves2008}, the slope of the colour-flux correlation can be used to infer the origin of the variability. The colour-brightness correlations shown in Fig.\,\ref{Fig_KK45} have in both cases a slope that agrees within the errors with the extinction law from \cite{Cardelli1989}. The extinction values at $3.6\mu$m and $4.5\mu$m for R\,=\,3.1 were taken from \cite{Chapman2009}. These clear correlations indicate that the variability is probably caused by variations in the extinction. The lack of any colour magnitude correlation for $\lambda$\,$\lesssim$\,2\,$\mu$m indicates that the absorbing dust particles have typical sizes of 1-2\,$\mu$\,m and are therefore larger than the typical dust particles in the interstellar medium. Fig.\,\ref{Fig_colourI} only suggests a weak tendency for CHS\,7797 to becomes bluer when fainter for $\lambda$\,$\lesssim$\,2\,$\mu$m, which if real, might be due to scattered light. 
\begin{table}[h!]
\centering
\begin{minipage}[c]{\columnwidth}
\renewcommand{\arraystretch}{1.6}  
\caption{Relevant properties of CHS\,7797.}
\label{properties}
\addtolength{\tabcolsep}{0.0pt}
\resizebox{8.8cm}{!} {
{\Large
\begin{tabular}{l|c|c} 
\hline\hline             
Parameter & \multicolumn{2}{c}{Value}\\
\hline
Position(2000) & \multicolumn{2}{c}{RA=05 35 08.7 DEC= -05 31 27.3}\\
\hline
Photometry\,$^{a}$ & Maximum (JD=2455137)& Minimum (JD=2455130)\\
I & 17.35&  19.01\\
R-I\,$^{b}$&2.25& 1.9 \\
I-J& 2.24& 2.25\\
J-K& 1.4 & 1.5\\
K$_{s}$-3.6& 1.3& 1.8\\
K$_{s}$-4.5& 2.2& 3.1\\
\hline
Typical ptp amplitudes\,$^{c}$& \multicolumn{2}{c}{I=1.7 z$\arcmin$=1.7 J=1.6 K=1.4 [3.6]=0.9 [4.5]=0.6}\\
\hline
Spectral type& \multicolumn{2}{c}{M6\,$^{d}$}\\
\textit{$T_{eff}$}& \multicolumn{2}{c}{2840-3000}\\
Distance\,(pc)&\multicolumn{2}{c}{450\,$^{e}$}\\
Luminosity\,(L$_{\odot}$)& \multicolumn{2}{c}{0.01-0.06\,$^{f}$}\\
Mass\,(M$_{\odot}$)&\multicolumn{2}{c}{0.055-0.1\,$^{f}$}\\
Age\,(Myr)&\multicolumn{2}{c}{1\,$^{e}$}\\
\hline
\end{tabular}}
}
\renewcommand{\footnoterule}{} 
\\
\\
$^{a}$ when available, the quasi-simultaneous photometric observations were used for the typical photometric values, since as seen in Fig.\,\ref{Fig_lightcurves1} there are variations in the minimum/maximum levels.\\
$^{b}$ the JD maximum is 2455474 and minimum 2455468 in this case.\\
$^{c}$ typical R-band ptp amplitudes are not listed because CHS\,7797 could not be reliably measured at minimum brightness.\\
$^{d}$ from Rodriguez-Ledesma et al 2012, in prep.\\
$^{e}$ from \cite{Odell2001}\\
$^{f}$ from this work.
\end{minipage}
\end{table}

\subsection{Stellar and disc properties of CHS\,7797\label{Luminosity}}

We followed several approaches for estimating the luminosity and mass of CHS\,7797.  We used the fluxes at maximum light corresponding to JD 2455137 (see the maximum peak in Fig.\,\ref{Fig_lightcurves3} for which I, J, K$_{s}$, IRAC1 and IRAC2 observations are available). We also considered that CHS\,7797 might be a binary system composed of identical spectral type (SpT) components (see Section\,\ref{Discussion}). We point out that the classical approach based on the dereddened J flux has to be considered with extreme caution since we are not dealing with normal interstellar dust, but with larger dust particles. Therefore we focused our luminosity and mass estimates on methods that are not, or only slightly, dependent on extinction. 

From a preliminary spectral analysis, and based on four narrow-band spectral indices particularly selected to measure the change of strength of molecular absorption feautures with spectral type, we derive a SpT of M6$\pm$\,0.5 for CHS\,7797  (for details see Rodriguez-Ledesma et al 2012, in prep.). We used the SpT-T$_{eff}$ relations derived by \cite{Luhman2003}, which is best suited for young objects. For an M6 young dwarf, the T$_{eff}$ based on this relation is 2990\,K. If the commonly used SpT-T$_{eff}$ relation for field dwarfs is used, one derives for CHS\,7797 a T$_{eff}$ of 2840\,K \citep{Luhman1999}. Based on the models of \cite{Baraffe1998} at 1\,Myr, the corresponding luminosities and masses for T$_{eff}$\,=\,2840/2990\,K are L\,=\,0.025/0.065\,L$_{\odot}$ and M\,=\,0.055/0.1\,M$_{\odot}$. It is important to note that the SpT-T$_{eff}$ relations for very young objects are relatively uncertain, and at these ages the available models are still relatively inaccurate. Nevertheless, CHS\,7797 probably has a mass and luminosity within the previously mentioned values. If CHS\,7797 is a binary system with identical spectral type components, we expect that each of the stars has a mass between 0.055 and 0.1\,M$_{\odot}$.

As a second approach, we used the available VOSA SED analyser \citep{Bayo2008} to compute the best set of parameters from the SED at maximum light. VOSA performs a statistical test to decide which set of synthetic photometry best reproduces the data. During the fitting process, the tool detects possible infrared excesses and does not consider those data points of the SED where the excess has been found, i.e. in our case the IRAC bands are not taken into account in the fitting. The best-fit corresponds to Teff=2700 log g=4 in the NextGen models, with a $\chi^2$ of 2. The derived bolometric luminosity is L\,=\,0.024\,L$_{\odot}$ if A$_v$=0 is considered and L\,=\,0.040\,L$_{\odot}$ if we assume an A$_v$=2.4\footnote{ This value is derived by using intrinsic (I-J) colours from \cite{Leggett1992} at maximum brightness. Nevertheless, we emphasise that the derivation of an A$_v$ has to be taken with caution since we are dealing with grey dust at $\lambda$\,$\lesssim$\,2\,$\mu$m.}. This luminosity is the luminosity of the system. Assuming a binary system, the \cite{Baraffe1998} models predict a mass of each component of 0.04 - 0.055\,$M_{\odot}$, in agreement with the derived masses and luminosities based on the SpT-T$_{eff}$ scale.

In addition to VOSA SED analyser, we made use of the SED fitting tool described in \cite{Robitaille2007}. This SED fitting tool uses the 200,000 model SEDs for young stellar objects presented in \cite{Robitaille2006}. The models are characterised by 16 stellar, envelope, and disc parameters such as stellar mass and temperature, the envelope accretion rate, disc mass, accretion rate, and inner disc radius. To infer some physical properties of CHS\,7797 and its disc, we compared the modelled SEDs with the observed SED taking into account the ten best-fitted models. The derived temperature of the central source ranges from 2700 to 3000\,K and the mass between 0.1 and 0.18\,$M_{\odot}$, for the different best-fitted models. It is important to note that the minimum mass considered in the models is 0.1\,$M_{\odot}$. All models suggest that the inclination angle of the system is between 75 and 82 degrees, and therefore we are seeing the system almost edge-on. As stated above, in addition to the stellar parameters these models allow for the characterisation of the circumstellar disc. The derived disc mass for CHS\,7797 ranges from $1x10^{-4}$ to $5x10^{-3}$\,$M_{\odot}$. 

As shown in Fig.\,\ref{Fig_SED} and also as computed with the VOSA SED analyser, the NIR excess starts at about 3$\mu$m, which suggests the presence of a gap in the inner disc. Based on the SED fitting tool, the inner disc edge in CHS\,7797 is most probably located between 0.2 and 0.6\,AU. The total range is larger (minimum of 0.04 and maximum at 2\,AU) but most of the derived values are between 0.2 and 0.6\,AU. It is interesting to compare the derived inner disc edge with the radius at which the Keplerian period corresponds to 17.8\,d ($R_{period}$). Assuming stellar masses of 0.07 and 0.2\,$M_{\odot}$, $R_{period}$ is 0.05 and 0.075\,AU, respectively. These values are much lower than the typical inner disc edge values derived. Only one of the best-fitted models suggests that the inner disc edge is slightly smaller than $R_{period}$. All other nine models indicate a much larger inner disc edge (up to 50 times larger than $R_{period}$), and therefore suggest the existence of an inner disc gap.

To summarise, we find that CHS\,7797 is a very low mass object whether it is a single or binary star, with masses that are probably substellar or close to the substellar boundary. In addition, there are indications of an inner disc gap.\\

\section{Discussion \label{Discussion}}

\subsection{The nature of the ''internal clock''}

The variability observed in CHS\,7797 is undoubtedly periodic with a period of 17.8 d. The periodic flux variations in young stars are usually interpreted by rotational modulation attributed to starspots in their atmospheres. The typical light-curves produced by these spots show quasi-sinusoidal shapes. In addition to a rotational motion, orbital motions are intrinsically periodic and can of course give rise to a periodic flux variation. As outlined in the introduction, this is true not only in eclipsing binary systems, but can also be caused by periodically variable absorption in a circumstellar/binary disc.

The 17.8\,d period measured for CHS\,7797 is rather long compared to the typical rotational periods measured in young very low mass objects. \cite{Rodriguez-Ledesma2009} found a median rotational period of $\approx$\,2\,d for 139 objects with estimated masses $\lesssim$\,0.1\,$M_{\odot}$ in the ONC. In the same work no objects with rotational periods $\gtrsim$\,13\,d were reported at these low masses. Studies in other young clusters reveal similar values for the rotational periods \citep[e.g.][found rotational periods of 2-3\,d for brown dwarf candidates in Cha\,I]{Joergens2003}.

Since cool spots can only produce brightness modulations I\,$<$\,0.7\,mag \citep{Herbst1994}, the very large amplitudes of I$\approx$\,1.7  observed in CHS\,7797 can only be caused by rotational modulation if we are dealing with a hot spot. If the light modulation of CHS\,7797 were caused by a hot spot, Fig.\,\ref{Fig_colourI} should show a correlation between brightness and colour, i.e. the star should become redder when becoming fainter. This is not observed. In addition, the variation amplitudes caused by hot spots must be larger in the optical than in the NIR, particularly for a cool object like CHS\,7797, because the relative contribution from the cool stellar atmosphere is strongly increasing towards the NIR, while the relative contribution from the hot spot is correspondingly decreasing. This is not observed for CHS\,7797, since the amplitudes do not decrease with wavelength for $\lambda$\,$\lesssim$\,1\,$\mu$m, and only slightly for $\lambda$\,$\gtrsim$\,1\,$\mu$m. Following the same approach as \cite{Carpenter2001}, we investigated the amplitudes of the brightness modulation that can be theoretically produced by hot and cool spots. We fixed the stellar T$_{eff}$ to 3000K, and assumed that spots can be characterised by single-temperature blackbodies covering a certain fraction \textit{f} of the photosphere. We investigated the amplitudes at R, I, J, K$_{s}$, $3.6\mu$m, and $4.5\mu$m, for hot spot temperatures of 5000\,K, 6000\,K, 7000\,K, 8000\,K, 9000\,K, and 10\,000\,K and for hot spots covering 1$\%$, 5$\%$, 10$\%$, 20$\%$, and 30$\%$ of the stellar surface. We find that to produce the amplitudes observed in CHS\,7797 at $\lambda$\,$\gtrsim$\,1\,$\mu$m ($\Delta$J\,=\,1.6 ,$\Delta$K$_{s}$\,=\,1.4 ,$\Delta$$3.6\mu$m\,=\,0.9 ,$\Delta$$4.5\mu$m\,=\,0.6), we need very hot spots (T$\approx$\,10\,000\,K) and spotted areas covering more than 20$\%$ of the stellar photosphere. These bright hot spots covering such a large part of the stellar surface would produce amplitudes in R and I bands of $\approx$\,4 and $\approx$\,3 mag, respectively, which are much larger than the $\approx$\,1.7\,mag observed in CHS\,7797. In addition, variability studies of TTS with hot spots often show phase jumps and substantial changes in their modulation amplitudes from epoch to epoch \citep[e.g.][]{Herbst1994}, while we observed only a moderate change in amplitude over the six years. We performed the same test for cool spot temperatures of 1500\,K, 1700\,K, 1900\,K, 2100\,K, 2300\,K, 2500\,K, and 2700\,K (i.e. 10\% to 50\% cooler than the photosphere) and for cool spots covering fractions from 10$\%$ to 75$\%$ of the stellar surface. We found that not even the coolest 1500\,K spot covering 75\% of the surface of CHS\,7797 can produce the large amplitudes observed in R, I, z, J, and K bands.  

We conclude from the previous arguments that a rotational light modulation due to hot or cool spots cannot explain the periodic variability of CHS\,7797, and therefore we most probably are dealing with an orbital motion.

\subsection{Nature of the variability}

The most likely source of orbital variability in an object such as CHS\,7797 is variable extinction from circumstellar matter, most probably located in a disc.

Our multiwavelength observations demonstrate that CHS\,7797 shows no correlation between colour and flux for $\lambda$\,$\lesssim$\,2\,$\mu$m. The NIR monitoring by \cite{Carpenter2001} also clearly indicates no colour-flux correlation in the NIR J,H,K$_{s}$ bands for CHS\,7797. Interestingly, we found a strong colour-flux correlation at $\lambda$\,$\gtrsim$\,2\,$\mu$m that agrees with variability caused by variable extinction (see Fig.\,\ref{Fig_KK45}). The lack of a colour-magnitude correlation at $\lambda$\,$\lesssim$\,2\,$\mu$m suggests that grain-growth has happened among the occulting material. Grain sizes with a minimum size of about 1 - 2\,$\mu$m in the circumstellar environment can account for the grey absorption at $\lambda$\,$\lesssim$\,2\,$\mu$m, while the distribution of grain sizes $\gtrsim$\,2\,$\mu$m should be similar to the interstellar distribution to explain the colour-flux correlation found at longer wavelengths, as clearly seen in Fig.\,\ref{Fig_KK45}. We note that various studies in the last decades showed that discs in TTS usually undergo grain processing \citep[e.g][]{Przygodda2003, Kessler2005, Shuping2006}. Based on 8 - 13\,$\mu$m low-resolution spectra of eight proplyds in the ONC, \cite{Shuping2006} found that the typical grain size in the surface of the discs is a few microns, similar to what was found in young TTS in other star-forming regions.

\subsection{Is the absorbing dust located in a circumstellar or circumbinary disc?}

CHS\,7797 could be a single object with a circumstellar disc and an structure or protoplanet orbiting with a period of 17.8\,d or it could be a low-mass stellar or brown dwarf binary system similar to KH\,15D.

\subsubsection{Single-object scenario}

In the single-object scenario, the eclipses in CHS\,7797 are caused by variable amounts of circumstellar dust in the line-of-sight to the object. Since CHS\,7797 is at the maximum brightness state only for $\approx$\,1/3 of the orbital phase, the structure in the circumstellar disc causing the ocultation has to be quite extended. This structure in the circumstellar disc is expected to orbit CHS\,7797 with a period of 17.8\,d. This scenario can only be taken into account if there is no inner disc cavity comparable in size with $R_{period}$. A large inner disc gap, as suggested by nine of the ten best-fitted SED models, would make the single model scenario rather unlikely. Nevertheless, since one of the ten best-fitted SED models indicates that the inner disc edge is $\lesssim$\,$R_{period}$, the single object scenario cannot be completely excluded. 

In the following, on the basis of the observations of V718\,Per \citep{Grinin2006,Grinin2008,Herbst2010b}, we argue that single periodically variable PMS stars with large variation amplitudes can be explained by strong absorption in their circumstellar discs over large fractions of their orbit.

The TTS V718\,Per is located in the 3Myr-old cluster IC\,348. It shows periodic variability by 0.8\,mag in I with a period of 4.7\,yr and eclipses that last $\approx$\,3/4 of the orbit \citep{Grinin2006,Grinin2008}. For comparison, the eclipses in CHS\,7797 last 2/3 of the orbit. The light-curve of V718\,Per \citep[see Fig.\,1 in][]{Grinin2008} looks smoother than the light-curve of CHS\,7797, but they are similar in shape. Although in the first attempts to explain its unusual periodic variability a binary system similar to KH\,15D was considered (see also next section), radial velocity measurements by \cite{Grinin2008} exclude the binary model for V718\,Per and put an upper limit to any companion to $\approx$\,6\,M$_{jup}$. The authors concluded that the star is surrounded by a nearly edge-on circumstellar disc with an irregular mass distribution presumably at the inner disc edge. As pointed out in their work, the extended circumstellar feature could be an irregular azimuthal mass distribution in the disc, such as spiral arms due to the interaction of the disc with a planet/protoplanet, or a warped disc structure. \cite{Herbst2010b} considered a similar model in which the disc structure is strongly altered by a massive planet/protoplanet. 

An additional example of this kind of variability is presented in \cite{Scholz2009}. Based on a NIR monitoring of 30 low mass objects in $\sigma$\,Ori (3\,Myr), \cite{Scholz2009} found evidence of a very low mass object (i.e. star $\#$2 in their work) in which variability might be caused by inhomogeneities at the inner edge of the disc. In all cases, the system has to be seen almost edge-on to account for the extinction variations. 

AA Tau-like stars also show variability due to inner disc perturbations that are associated with a magnetic interaction between the star and the disc \citep{Alencar2010}. AA Tau-like light curves are found to be fairly common among young stars, but the comparison with the light curves of CHS\,7797 reveals significant differences. First, AA Tau-like stars are only quasi-periodic, (i.e. show time evolution over a few cycles) and not at all as perfectly periodic as CHS\,7797 (\textit{FAP}\,=\,10$^{-15}$). Second, the light curves of AA Tau systems have a flat maximum that is interrupted by minima events that are highly variable both in width and depth, quite different from the shape of the light curves of CHS\,7797, as shown in Fig.\ref{Fig_lightcurves1} and \ref{Fig_phasedlc_season}.

\subsubsection{Interpreting our data in terms of a KH\,15D type binary model}

In the binary scenario, CHS\,7797 is a low-mass stellar or brown dwarf binary system with an orbital period of 17.8\,d and a circumbinary disc. The orbital period of 17.8\,d and the estimated separation between the binary components of $\approx$\,0.08\,AU (for M$_{1}$\,=\,M$_{2}$\,=\,0.1\,M$_{\odot}$) are consistent with other known young low-mass binaries \citep[e.g.][]{Stassun2006}. By analogy with KH\,15D, we assumed that the disc is slightly inclined with respect to the orbital plane of the binary and that \textit{only one} of the components can be seen during apastron through a more transparent outer region of the disc during $\approx$\,1/3 of the orbit. This means that -- in contrast to KH\,15D, for which until 2007 one star (star A) was unocculted by the circumbinary disc during maximum brightness -- CHS\,7797 is always occulted by the disc but the amount of extinction is periodically changing depending on the position of one of the stars with respect to the more transparent outer regions of the circumbinary disc. Note that the $\approx$\,1.7 peak-to-peak amplitude of CHS\,7797 in the I band is much smaller than the nearly 4 mag amplitude observed in KH\,15D. One possibility to account for the variations at maximum level in CHS\,7797 and the general variability in I-band of typically 0.3\,mag at a given phase, is a disc that does not extend smoothly and homogeneously to its outer edge but that is highly structured, and therefore we see the star at the same orbital phase through portions of the disc that are correspondingly variable in transparency (in some cases from one orbital period to the following one). These variations in transparency can in principle also explain the ``supermaxima'' that we observed in our data (see Fig.\,\ref{Fig_lightcurves1}), in which the object is about 1\,mag brighter in I band than the typical maxima level. Over the six years of monitoring, only two supermaxima were detected (an additional one with large photometric errors was ignored), which clearly indicates that these extreme increases in brightness are rare. We note that the ``supermaxima'' events do not occur at maximum phase, but some days after maximum ($\approx$ at the end of ingress phase). Another possible explanation for the observed ``supermaxima'' events is flaring activity, which cannot be excluded for CHS\,7797 because it is a strong H$\alpha$ emitter (Rodriguez-Ledesma et al., in prep.). For the same reason the $\approx$\,0.3\,mag variations can be accretion-related. We expect the latter effect to contribute only in $\approx$\,0.05\,mag colour changes in the (R-I) versus I diagram (Fig.\,\ref{Fig_colourI}), and we note that some CTTSs show little colour-brightness correlations \citep[e.g. AA\,Tau, DR\,Tau][]{Herbst1994}.

The present interpretation of the KH\,15D system considers the precession of the circumbinary disc, which causes the occulting sharp edge to move across the orbit of the binary \citep{Winn2006, Herbst2010}. This substantial difference can be appreciated if one compares the light-curves of CHS\,7797 in Fig.\,\ref{Fig_lightcurves1} with the light-curves of KH\,15D \citep[see e.g. Fig.\,3 in][]{Herbst2010}: KH\,15D light-curves have a constant maximum brightness within the same observing season, and only show a steady and smooth decrease in amplitude with years. Furthermore, the shape of the light-curves in KH\,15D also changes in a continuous and smooth fashion from one observing season to another. As seen in Fig.\,3 of \cite{Herbst2010}, the star is at bright state for a shorter amount of time every consecutive observing season. Over the six years of monitoring of CHS\,7797 we did not observe such a monotonic decrease (or increase) of the maximum light duration. This implies that any possible disc precession in CHS\,7797 occurs at a much lower rate than in KH\,15D. We note that the typical shape of the light-curve brightness modulation shown by CHS\,7797 is similar to that shown by KH\,15D during the 2007/2008 season \citep[Fig.\,3 in][]{Herbst2010}. 

\section{Conclusions}

CHS\,7797 is undoubtedly a very interesting and unusual young and possibly substellar object that is occulted by circumstellar/circumbinary matter. Whether it is a single object or a binary system, CHS\,7797 is surrounded by a circumstellar/circumbinary disc that is viewed almost edge-on and in which substantial grain growth has taken place. A comparison of the light-curves and phased light-curves of CHS\,7797 with the three other known ``unusual`` periodic variables (i.e. KH\,15D, WL\,4 and V718\,Per) suggests some similarities but also discrepancies. In particular, the other three objects show much smoother light curves than CHS\,7797, which implies that, for as yet unknown reasons, the disc around CHS\,7797 is much more inhomogeneous, and/or that there is a contribution of accretion-related irregular variations. Clearly, CHS\,7797 deserves further attention. An upcoming spectroscopic analysis (Rodriguez-Ledesma et al., in prep) will try to provide additional constraints on the nature of CHS\,7797. Nevertheless, since CHS\,7797 is very faint and since there is a strong background contribution from the Orion nebula, it will be hard to search for radial velocity variations to confirm the possible binary character of the system even with the instrumentation currently available at 8-10\,m class telescopes.

\begin{acknowledgements}
The authors would like to thank all observers at MMO and the 2.2m/WFI, at La Silla, Chile, for their support on the observation. We thank the anonymous referee, whose comments helped to improve the paper. This publication makes use of VOSA, developed under the Spanish Virtual Observatory project supported from the Spanish MICINN through grant AyA2008-02156. This research was partially funded by the German Research Foundation (DFG) grant HE\,2296/16-1.
\end{acknowledgements}
\bibliographystyle{aa}
\bibliography{chs7797a}

\appendix
\section{Absolute photometry of stars in the CHS\,7797 region}
We have performed Bessell R, I, and SDSS z$\arcmin$ absolute photometry of nine stars in the vicinity of CHS\,7797 based on observations at MMO. These stars were used as secondary standards for the relative photometry described in Section 3 and have been selected from the work of \cite{Rodriguez-Ledesma2009} as non-variable (or very low variable) stars. Although the ONC is a well-studied region, due to the bright and inhomogeneous nebular background no absolute R and z$\arcmin$ photometry was available so far. We therefore performed the absolute photometry by observing Landolt standard fields and Sloan Sky Digital Survey (SSDS) fields. The table below summarises the results.
\begin{table}[h!]
\centering
\begin{minipage}[c]{\columnwidth}
\renewcommand{\arraystretch}{1.5}  
\caption{MMO R, I, z$\arcmin$ photometry for ONC stars in the vicinity of CHS\,7797.
}
\label{catalog}
\addtolength{\tabcolsep}{0.0pt}
\resizebox{8.8cm}{!}{
\begin{tabular}{l|c|c|c|c|c|c} 
\hline\hline             
ID$^{a}$ & R&R$_{err}$&I&I$_{err}$&z$\arcmin$&z$\arcmin_{err}$\\
\hline
255 & 14.643 & 0.071 & 14.019 & 0.071 & 15.364 & 0.084 \\
209 & 15.748 & 0.071 & 14.315 & 0.071 & 14.727 & 0.084 \\
10318 & 15.557 & 0.100 & 14.285 & 0.095 & 14.798 & 0.110 \\
148 & 16.933 & 0.089 & 15.304 & 0.082 & 14.923 & 0.105 \\
105 & 18.258 & 0.145 & 16.015 & 0.108 & 15.144 & 0.122 \\
120 & 16.304 & 0.071 & 14.572 & 0.070 & 14.524 & 0.084 \\
132 & 16.342 & 0.077 & 14.706 & 0.071 & 14.734 & 0.084 \\
10263 & 18.577 & 0.158 & 17.028 & 0.102 & 17.269 & 0.207 \\
103 & 18.432 & 0.138 & 16.350 & 0.088 & 15.553 & 0.130 \\  
\hline
\end{tabular}
}
\renewcommand{\footnoterule}{} 
\\
\\
$^{a}$ ID numbers are from \cite{Hillenb1997}.  
\end{minipage}
\end{table}

\end{document}